\documentstyle[aps,prd,tighten,amsmath,epsfig]{revtex}

\begin{document}
\draft
\twocolumn

\widetext

\title{Neutrino oscillations and mixings with three flavors}
\author{Tommy Ohlsson\footnote{Electronic address:
tommy@theophys.kth.se} and H{\aa}kan Snellman\footnote{Electronic address:
snell@theophys.kth.se}}
\address{Theoretical Physics, Department of
Physics, Royal Institute of Technology, SE-100 44 Stockholm, Sweden}
\date{Received 4 March 1999}

\maketitle

\begin{abstract}
Global fits to all data of candidates for neutrino oscillations
are presented in the framework of a three-flavor model. The analysis
excludes mass regions where the MSW effect is important for the solar
neutrino problem. The best fit gives $\theta _{1} \approx 28.9^\circ$,
$\theta_{2} \approx 4.2^\circ$, $\theta_{3} \approx 45.0^\circ$,
$m_{2}^{2}-m_{1}^{2} \approx 2.87 \times 10^{-4} \; {\rm eV}^2$, and
$m_{3}^{2}-m_{2}^{2} \approx 1.11 \; {\rm eV}^2$ indicating essentially maximal
mixing between the two lightest neutrino mass eigenstates.
\end{abstract}


\section{Introduction}
\label{sec:intro}

Neutrino oscillations were reported last year by the
Super-Kamiokande Collaboration \cite{kaji98} to have been convincingly seen
by the Super-Kamiokande detector in Kamioka, Japan, in the study of
the atmospheric neutrino problem. This has rekindled interest to
explain also other deviations from expected results in neutrino
physics in terms of neutrino oscillations.
There are at present essentially four different types of experiments
looking for neutrino oscillations: solar neutrino experiments, atmospheric
neutrino experiments, accelerator neutrino experiments, and reactor neutrino
experiments.

Most analyses of neutrino data so far have been performed in
two-flavor models. Although this is quite illustrative, it can lead to
wrong conclusions concerning the necessary number of degrees of
freedom to explain data. It is, e.g., possible that the four
neutrino scenario, advocated by several authors, is an artifact of this
simplification.

Recently, several papers
\cite{thun98,bare98,conf98,tesh98,barg98,saka99,acke97,fogl99} 
have been published pertaining at demonstrating the possibility to obtain an
overall description in terms of a three-flavor model. Most of these
investigations have focused on demonstrating the plausibility of the
scenario, without truly fitting all the data.

Here we would like to report on the first stage of a global
analysis of all the neutrino oscillation data in a three-flavor model.
We will assume that $CP$ nonconservation is negligible at the present
level of experimental accuracy.
Thus, the Cabibbo-Kobayashi-Maskawa (CKM) mixing matrix for the
neutrinos is real.
There are then in principle five parameters that can be fitted in this
model: Three mixing angles and two mass squared differences. (This is
in contrast with two two-flavor models, which give only four parameters:
Two mixing angles and two mass squared differences.) The two mass
squared differences in the three-flavor model enter the argument of
the sinodial function in combination with the ratio $L/E$, where $L$
is the source-detector distance and $E$ is the neutrino energy.  The
argument of the sinodial functions related to the
energy dependence can then be treated as yet another angle.  In this case,
the probability function can be seen as given by a rotation of the flavor
states by five angles.  This is in analogy to the two-flavor case, which can
be seen as a rotation by two angles, one of which is given by $\Delta
m^{2}L/4E$, $\Delta m^{2}$ being the mass squared difference of the two
neutrino masses.

In the present analysis, we will make the simplifying assumption that
the relevant ranges of mass squared differences are such that the
Mikheyev-Smirnov-Wolfenstein (MSW) effect \cite{mikh85} in the Sun is
negligible. This excludes essentially the sensitivity region
$10^{-6}$--$10^{-5}$ eV$^{2}$.
The absence of any day-night variation in the Super-Kamiokande and
Kamiokande data is also consistent with neglecting the MSW resonance
effect in the Earth for the electron-neutrino data.

The differing results for the solar neutrino experiments present another
problem. Since the Super-Kamiokande and Kamiokande experiments can
resolve the energy
for the solar neutrinos, these experiments are of a rather different and
more detailed kind than the radio-chemical experiments.  It is possible to
vary the disappearance rate by going to a mass squared realm, where the
arguments, i.e., the angles that depend on the mass squared differences,
are sensitive to the energy of the neutrinos, since this is different for
the different detection methods.  However, in this case, the energy domains
for the different experiments do not vary in the way expected, making the
Cl experiment the smallest ($\approx 0.33$) and the Ga experiments the
largest ($\approx 0.6$) of the expected rate.  One would, in fact,
expect the Super-Kamiokande
experiment, with the highest mean energy for the neutrinos, to give the
smallest result, whereas in reality it is in between.

Confronting this with the apparent lack of variation in the
experimental solar neutrino detection probability with energy, at
least up to approximately 13 MeV, in the Super-Kamiokande data, we
will adopt the point of view that the mass squared differences are in
a range that leads to the solar neutrino oscillations being averaged
out to their classical values. By fitting the three mixing angles, we
then account for the weighted mean of all the solar neutrino
experiments.  By nature of the sensitivity level of the
accelerator and atmospheric neutrino experiments, we will choose the mass
squared difference ranges as $0.2 \; {\rm eV^2} \leq \Delta M^{2} \leq 2
\; {\rm eV^2}$ and $10^{-4} \; {\rm eV^2} \leq \Delta m^{2} \leq 10^{-3}$
eV$^{2}$.  Here $\Delta M^{2}$ will regulate the accelerator experiments
and the multi-GeV atmospheric neutrino data,
whereas $\Delta m^{2}$ regulates the low energy atmospheric experiments.
This choice is thus not inherited from two-flavor model analyses, but
arises from the sensitivity of the different types of experiments.

It is in principle possible that there would be a resonance effect
for the atmospheric electron-neutrinos passing through Earth. Our
chosen ranges for the mass squared differences exclude this effect in
the present fit, since $\Delta m^2 \geq 10^{-4}$ eV$^{2}$.

Since the reactor experiments are disappearance experiments with no signal
seen, they can at best exclude certain ranges for the mass squared
differences due to their sensitivity level.  The two ranges of mass squared
differences chosen here respect this sensitivity level.

The paper is organized as follows.
In Sec.~\ref{sec:form}, we briefly discuss the formalism used in our
analysis including notations, Gaussian averaging, and minimization. In
Sec.~\ref{sec:CoD}, we discuss the choice of data from different
experiments used in the
minimization procedure. Sec.~\ref{sec:NA} describes the minimization
procedure and includes the results thereof. We also show that our results are
consistent with the experimental data, assuming that all of the
deviations from the expected values are indeed related to the physics
of neutrino oscillations. Finally, Sec.~\ref{sec:DC} presents a
discussion of the results and also our main conclusions.

\section{Formalism}
\label{sec:form}

\subsection{Notations}
\label{sub:not}

In the present analysis, we will use the plane-wave approximation to
describe neutrino oscillations. In this approximation, a neutrino
state $\vert \nu_\alpha \rangle$ with flavor $\alpha$ is a linear
combination of neutrino mass eigenstates $\vert \nu_a \rangle$ such that
\begin{equation}
\vert \nu_\alpha \rangle = \sum_{a=1}^3 U_{\alpha a}^\ast \vert \nu_a
\rangle, \quad \alpha = e, \mu, \tau,
\end{equation}
where the $U_{\alpha a}$'s are entries in a unitary $3 \times 3$
matrix $U$.

The unitary matrix $U$ is given by
\begin{equation}
U = (U_{\alpha a}) = \left( \begin{array}{ccc} U_{e 1} & U_{e 2} &
U_{e 3} \\ U_{\mu 1} & U_{\mu 2} & U_{\mu 3} \\ U_{\tau 1} & U_{\tau
2} & U_{\tau 3} \end{array} \right).
\end{equation}
A convenient parametrization for $U$ is \cite{caso98}
\begin{equation}
U = \left( \begin{array}{ccc} c_2 c_3 & s_3 c_2 & s_2 \\ - s_3 c_1 -
s_1 c_3 s_2 & c_1 c_3 - s_1 s_2 s_3 & s_1 c_2 \\ s_1 s_3 - s_2
c_1 c_3 & - s_1 c_3 - s_2 s_3 c_1 & c_1 c_2 \end{array} \right),
\end{equation}
where $s_i \equiv \sin \theta_i$ and $c_i \equiv \cos \theta_i$ for $i
= 1,2,3$. This is the so called standard representation of the
CKM mixing matrix. We have here put the $CP$
phase equal to zero in the CKM matrix. This means that $U^\ast_{\alpha
a} = U_{\alpha a}$ for $\alpha = e, \mu, \tau$ and $a = 1,2,3$.

The probability of transition from a neutrino flavor state $\alpha$ to
a neutrino flavor state $\beta$ is given by
\begin{eqnarray}
P_{\alpha\beta} &\equiv& P(\nu_\alpha \to \nu_\beta) \nonumber \\
&=& \delta_{\alpha\beta} - 4 \; \underset{a < b}{\sum_{a=1}^3
\sum_{b=1}^3} U_{\alpha a} U_{\beta a}
U_{\alpha b} U_{\beta b} \nonumber \\
&\times& \sin^2 \frac{\Delta m_{ab}^2 L}{4E}, \quad
\alpha,\beta = e, \mu, \tau,
\label{eq:P_ab}
\end{eqnarray}
where $\delta_{\alpha\beta}$ is Kronecker's delta and $\Delta m_{ab}^2
\equiv m_a^2 - m_b^2$. The three mass squared differences $\Delta
m_{21}^2$, $\Delta m_{32}^2$,
and $\Delta m_{13}^2$ are not linearly independent, since they
satisfy the relation
\begin{equation}
\Delta m_{21}^2 + \Delta m_{32}^2 + \Delta m_{13}^2 = 0.
\end{equation}
Therefore, we will define
\begin{eqnarray}
\Delta m^2 &\equiv& \Delta m_{21}^2 = m_2^2 - m_1^2, \\
\Delta M^2 &\equiv& \Delta m_{32}^2 = m_3^2 - m_2^2,
\end{eqnarray}
and thus $\Delta m_{31}^2 = - \Delta m_{13}^2 = \Delta m^2 +
\Delta M^2$.

From Eq.~(\ref{eq:P_ab}) it is clear that $P({\bar{\nu}}_\beta \to
{\bar{\nu}}_\alpha) = P(\nu_\alpha \to \nu_\beta)$ and since we have
assumed that $CP$ is conserved, i.e., the CKM mixing matrix $U$ is
real, it holds that $P(\nu_\beta \to \nu_\alpha) = P(\nu_\alpha \to
\nu_\beta)$. This implies that
\begin{equation}
P({\bar{\nu}}_\alpha \to {\bar{\nu}}_\beta) = P(\nu_\alpha \to
\nu_\beta).
\end{equation}

\subsection{Gaussian averaging}
\label{sub:ga}

Since in practice the neutrino wave is neither detected nor produced
with sharp energy or with well defined propagation length, we have to average
over the $L/E$ dependence and other uncertainties in the detection and
emission of the neutrino wave.
We will here use the Gaussian average, which is defined by
\begin{equation}
\langle P \rangle = \int_{-\infty}^{\infty} P(x) f(x) \, dx,
\end{equation}
where
$$
f(x) = \frac{1}{\gamma \sqrt{2 \pi}} e^{-(x-l)^2/2 \gamma^2}.
$$
Here $l$ and $\gamma$ are the expectation value and standard
deviation, respectively.

By taking the Gaussian average of Eq.~(\ref{eq:P_ab}), we obtain the
average transition probabilities from flavor $\nu_\alpha$ to flavor
$\nu_\beta$ as
\begin{eqnarray}
\langle P_{\alpha\beta} \rangle &=& \delta_{\alpha\beta} - 2 \;
\underset{a < b}{\sum_{a=1}^3 \sum_{b=1}^3} U_{\alpha a} U_{\beta a}
U_{\alpha b} U_{\beta b} \nonumber \\
&\times& \left[ 1- \cos \left(2 l \Delta
m_{ab}^2\right) e^{-2 \gamma^2 (\Delta m_{ab}^2)^2} \right], \nonumber \\
&& \alpha,\beta = e, \mu, \tau.
\label{eq:<P_ab>}
\end{eqnarray}

The physical interpretations of the parameters $l$ and $\gamma$
are the following: The parameter $l$ deals with the
sensitivity of an experiment and is given by $l \simeq 1.27 \langle L/E
\rangle$, where $\langle L/E \rangle$ should be
given in km/GeV or m/MeV, and $\Delta m_{ab}^2$ should be measured in
${\rm eV}^2$. Note that we will here use $\langle L/E \rangle =
\langle L \rangle / \langle E \rangle$. The parameter
$\gamma$ is a so called damping factor. For large values of $\gamma$,
the dependence on the mass squared differences will be completely
washed out, since $1 - \cos \left(2 l \Delta m_{ab}^2\right) e^{-2
\gamma^2 (\Delta m_{ab}^2)^2} \to 1$ when $\gamma \to \infty$, and the
transition probabilities $\langle P_{\alpha \beta} \rangle$ will just
be dependent on the mixing angles $\theta_i$, where $i = 1,2,3$. This
corresponds to the classical limit. In the other limit, $\gamma \to 0$,
we will just regain Eq.~(\ref{eq:P_ab}) from Eq.~(\ref{eq:<P_ab>}),
i.e., $\lim_{\gamma \to 0} \langle P_{\alpha \beta} \rangle =
P_{\alpha \beta}$.

\subsection{Minimization}
\label{sub:mini}

In order to obtain the mixing angles $\theta_i$, where $i=1,2,3$, and
the mass squared differences $\Delta m^2$ and $\Delta M^2$, we
minimize the following object function
\begin{eqnarray}
F &\equiv& F(\theta_1,\theta_2,\theta_3,\Delta m^2, \Delta M^2)
\nonumber \\
&=& \frac{1}{w}
\sum_{i=1}^{N_{\rm exp}} w_i \frac{1}{{P^{\rm exp}_i}^2} \left( P_i -
P^{\rm exp}_i \right)^2,
\label{eq:obj}
\end{eqnarray}
where $w = \sum_{i=1}^{N_{\rm exp}} w_i$ and $N_{\rm exp}$ is the
number of neutrino oscillation experiments. The function $F$ is a
least square method error function for the relative errors of the
neutrino oscillation experiments with weights $w_i \geq 0$, where
$i=1,2,\ldots,N_{\rm exp}$. The quantities $P_i = \langle P_{\alpha
\beta} \rangle_i$ and $P^{\rm exp}_i$ are the theoretical (model) and
experimental transition probabilities for the $i$th experiment, respectively.
The reason why we are using relative errors in the error function and
not absolute errors (as is usually done) is due to the fact that
the different categories of experiments have rather different numerical
ranges for the experimental transition probabilities.
In the rest of this paper, we will write $P_{\alpha\beta}$, $L$, and
$E$ instead of $\langle P_{\alpha\beta} \rangle$, $\langle L \rangle$,
and $\langle E \rangle$, respectively, to keep the notations simpler.

In accordance with the discussion in the Introduction, the following
constraints will be used in the minimization procedure
\begin{eqnarray}
&& 0 \leq \theta_i \leq \frac{\pi}{2}, \quad \mbox{where $i=1,2,3,$}
\nonumber \\
&& 10^{-4} \; {\rm eV}^{2} \leq \Delta m^2 \leq 10^{-3} \; {\rm
eV}^{2}, \nonumber \\
&& 0.2 \; {\rm eV}^{2} \leq \Delta M^2 \leq 2 \; {\rm eV}^{2}. \nonumber
\end{eqnarray}
The weights in the object function~(\ref{eq:obj}) will be chosen as
\begin{equation}
w_i \sim \frac{1}{n \epsilon_i},
\end{equation}
where $n$ is the number of experiments used in the minimization of the
same category (solar, atmospheric, accelerator, or reactor) and
$\epsilon_i = \Delta P_i/P^{\rm exp}_i$ is the relative error in the
$i$th experiment.

Our choice for the weights is motivated by our wish to treat all
categories of experiments on the same footing (the factor $1/n$), and
enhance experiments with small relative errors (the factor
$1/\epsilon_i$). Here again, relative errors are used, since the range
of values of the probabilities are rather different. However, in all
our cases, $\epsilon_i < 1$. Higher powers of $1/\epsilon_i$ would
enhance the good experiments too much. Since $\Delta P_i$ is a sum of
the statistical and systematical errors, an underestimation of the
latter will, of course, also overemphasize the relative weight of a
certain experiment.

For the damping factor of an experiment, we will use
\begin{equation}
\gamma \sim \frac{L}{E} \left( \frac{\Delta L}{L} +
\frac{\Delta E}{E} \right),
\end{equation}
where $\Delta L$ is a combination of the uncertainties coming from (a)
the length of the detector and (b) the distance between the detector
and the neutrino source and $\Delta E$ is the uncertainty in the mean
neutrino energy.
In Table~\ref{tab:experiments}, we have listed all the experiments, which we
have used in the minimization procedure.

\section{Choice of Data}
\label{sec:CoD}

Most of the data used in our analysis are taken directly from the
quoted publications. (See Table~\ref{tab:experiments} for the
references to each experiment.) We have chosen in this first analysis
not to discriminate between the various experiments in any other way
than by their experimental uncertainties.

\subsection{Solar neutrino experiments}
\label{sub:sne}

For the chlorine (Cl) and gallium (Ga) experiments, we will define the mean
neutrino energy as
\begin{equation}
E = \frac{1}{S} \sum_r \langle E \rangle_r S^{(r)},
\end{equation}
where $S = \sum_r S^{(r)}$. The index $r$ labels the different solar neutrino
sources. In the case of a Cl experiment (Homestake), $r$ can be {\it pep},
$^7$Be, $^8$B, {\it hep}, $^{15}$O, or $^{17}$F, and similarly for a
Ga experiment (SAGE or GALLEX), $r$ can be {\it pp}, {\it pep},
$^7$Be, $^8$B, {\it hep}, $^{13}$N, $^{15}$O, or $^{17}$F. The
quantities $\langle E \rangle_r$ and $S^{(r)}$ are the average
neutrino energy and standard solar model (SSM) prediction for the
neutrino capture rates of the solar neutrino source $r$,
respectively. Using the values given by the
Bahcall-Basu-Pinsonneault 1998 (BBP98) SSM \cite{bahc98,bile98}, we obtain
the mean neutrino energy to be $E_{\rm Ga} \approx 1.1 \; {\rm MeV}$
for the SAGE and GALLEX experiments and $E_{\rm Cl} \approx 5.5 \;
{\rm MeV}$ for the Homestake experiment.

As for the water-Cherenkov detector experiments (Super-Kamiokande and
Kamiokande), we have chosen the mean energy of the neutrinos to be
$E_{\rm w-C} \approx 10 \; {\rm MeV}$.
We will assume that the uncertainty in the energy is $\Delta E
\sim E$, respecting the threshold energy for detection. For the
Super-Kamiokande and Kamiokande experiments, the energy resolution is
taken to be $\Delta E \sim 2$ MeV.

For solar neutrino experiments, the path length for the neutrinos is given by
the mean Sun-Earth distance. The length of the detector, the
deviation in the Earth's orbit around the Sun, and the solar radius
are all small in this connection, giving $\Delta L/L \sim 0$ for solar
neutrinos.

\subsection{Atmospheric neutrino experiments}

The path length of the atmospheric neutrinos is given by
\begin{equation}
L \equiv L(\vartheta) = \sqrt{R^2 \cos^2 \vartheta + 2 R d + d^2} - R \cos
\vartheta,
\end{equation}
where $\vartheta$ is the so called zenith angle, $R$ is the radius of
the Earth, and $d$ is the altitude of the production point of
atmospheric neutrinos. We have taken $R
\approx 6400 \; {\rm km}$ and $d \approx 10 \; {\rm km}$, which gives
$L \approx 1.22 \times 10^4 \; {\rm km}$ for the bin mean value $\cos
\vartheta = -0.95$ (MACRO and Kamiokande) and $L \approx 1.03 \times
10^4 \; {\rm km}$ for bin mean value $\cos \vartheta = -0.8$ (Soudan,
Super-Kamiokande, and IMB). The bins $-1 \leq \cos \vartheta \leq
-0.9$ and $-1 \leq \cos \vartheta \leq -0.6$ have $\Delta L/L \sim
0.10$ and $\Delta L/L \sim 0.50$, respectively.

The atmospheric neutrino experiments present a special
problem, since they are not entirely comparable to each other.
The Super-Kamiokande experiment measures several parameters and has presented
the ``ratio of ratios'' integrated over the
energy (actually $L/E$) as does also the Kamiokande experiment.
Other experiments can only measure the ratio of observed neutrino
flux and expected neutrino flux. The double ratio eliminates
most of the uncertainty in the muon-neutrino flux. However, this
function is also more complicated to fit,
and has lost much of the information of path length and energy of the
neutrinos. We have therefore used the directly measured
disappearance rates evaluated from the data for upward-going events given
in the tables and figures of
Refs. \cite{kaji98,saka99,gall98,ambr98,hata98,beck92}. This makes it
possible to use several of the other
atmospheric neutrino experiments on the same footing. The disadvantage
is the uncertainty in the expected neutrino flux. This uncertainty is
claimed to be of the order of $20\%$ \cite{kaji98}.

The main uncertainty of the neutrino flux comes from the uncertainty in
the primary cosmic radiation flux giving rise to the pions and kaons
responsible for the muon and (later) electron neutrinos. Let the
overall factor of flux uncertainty be
denoted by $\eta_{\alpha}$ for neutrino flavor $\nu_{\alpha}$. Thus,
e.g., the actual muon-neutrino flux is $\tilde{\Phi}_{\mu} =
\eta_{\mu}\Phi_{\mu}$, where $\Phi_{\mu}$ is the theoretical muon-neutrino
flux calculated in Ref. \cite{hond95}. Then, in the absence of
neutrino oscillations, the ratio of measured and theoretical flux is
just $R_{\mu}= \eta_{\mu}$.
In the presence of neutrino oscillations, on the other hand, we
instead obtain
\begin{equation}
R_{\alpha} = \frac{\tilde{\Phi}_{e}P_{e\alpha} +
\tilde{\Phi}_{\mu}P_{\mu\alpha} +
\tilde{\Phi}_{\tau}P_{\tau\alpha}}{\Phi_{\alpha}},
\end{equation}
for any flavor $\alpha = e, \mu, \tau$. Here $P_{\alpha\beta}$ is the
probability of oscillation from flavor $\alpha$ to flavor $\beta$.
Since the ratio $R = \Phi_{\mu}/\Phi_{e}$ is considered to be quite
well determined theoretically, perhaps even to $5\%$, due
to known branching ratios of the pions \cite{hond95} and their decays, we
conclude that the various factors $\eta_{\alpha}$ are essentially the
same for $\alpha = e,\mu$. This is consistent with the picture, which
says that the uncertainty is due to uncertainty in the primary
cosmic ray flux. We will further also neglect contributions from
the $\tau$ leptons in what follows. They should come from production
of $c$ or $b$ mesons, but the relevant cross sections are small
compared to the pion cross sections.

For the muon-neutrino ratio measured by several experiments, we obtain
\begin{equation}
R_{\mu} = \eta \left( P_{\mu\mu} + \frac{1}{R} P_{e\mu} \right).
\label{eq:rmu}
\end{equation}
At $E\approx 10$ GeV (Super-Kamiokande), we have $R \simeq 3$
\cite{kaji98,hond95}, and the last term can be
neglected if $P_{e\mu}$ is small, which turns out to be the case. Thus,
we obtain $P_{\mu\mu} \simeq \eta^{-1} R_\mu$. We will therefore calculate
the solution with the overall factor $\eta$ varying from $0.8$
to $1.2$ in the experimental values for $R_{\mu}$ of upward-going
events with $\cos \vartheta \leq - 0.6$.

For the electron-neutrino ratio $R_{e}=\eta(P_{ee}+RP_{\mu e})$, the
term $R P_{\mu e}$ is not negligable compared to the term $P_{ee}$,
and can therefore not be neglected. Instead of using $R_e$ as input, we
will analyze $R_e$ as an outcome of the optimization process,
i.e., in terms of the fitted parameters.

Finally, the sensitivity for the atmospheric neutrinos can be divided into two
groups. The first group (Soudan and Super-Kamiokande) is
sensitive to the small mass squared difference only, whereas the second group
(MACRO, Kamiokande, and IMB) is sensitive both to the large and
(slightly) to the small mass squared differences. Changing the ratio
from $0.8$ to $1.2$ times the quoted ones
(Table~\ref{tab:experiments}), will therefore influence the
determination of the mass squared differences.

\subsection{Accelerator neutrino experiments}
\label{sub:ane}
The mean energies and path lengths for the accelerator neutrino experiments
are all given in Refs. \cite{atha95,zeit98,alte98,migl98}. In most
cases, a measure of $\Delta L$ is given by half the length of the
detector, giving $\Delta L/L$ in the range $0.15$--$0.35$.
For the transition probabilities,
we have used the presented $90\%$ C.L. upper bounds, except for the
Liquid Scintillator Neutrino Detector (LSND)
experiment, for which we have used their quoted $\nu_{\mu}\rightarrow
\nu_{e}$ transition probability.

\subsection{Reactor neutrino experiments}
\label{sub:rne}
The reactor neutrino experiments are all disappearance experiments
with no signal detected. We have used the published $90\%$ C.L. bounds
on disappearance signals mainly to limit the mass squared ranges. Once
these are outside the sensitivities of the reactor experiments, the
solution will automatically reproduce these.  For the reactor experiments,
the uncertainties in $L$ are all small compared to the uncertainties in
energy determination.

\section{Numerical Analysis}
\label{sec:NA}

To be able to minimize Eq.~(\ref{eq:obj}) we have to use numerical
methods, since there is no successful way to do this with analytical
methods, the present function being to large and we have far too many
variables. Our procedure will be as follows. First, we will use a very
simple stochastic procedure. We choose some number ($n$) of random points
in the specified domain and evaluate the object
function~(\ref{eq:obj}). We simply take the point which gives the
smallest value of this function. To obtain good statistics we repeat
this procedure $N$ times. Including all 16 experiments, discussed in
the previous section, and choosing $n = 10^6$ and $N = 20$, we obtained
\begin{eqnarray}
&& \theta_1 = 28.7^\circ \pm 0.7^\circ, \nonumber \\
&& \theta_2 = 4.4^\circ \pm 0.1^\circ, \nonumber \\
&& \theta_3 = 45.1^\circ \pm 0.3^\circ, \nonumber \\
&& \Delta m^2 = (2.89 \pm 0.21) \times 10^{-4} \; {\rm eV}^2, \nonumber \\
&& \Delta M^2 = 1.12 \pm 0.04 \; {\rm eV}^2, \nonumber
\end{eqnarray}
where we have used
$$
x = \bar x \pm \frac{1}{\sqrt{N}} \sigma_{N-1}.
$$
The quantities $\bar x$ and $\sigma_{N-1}$ are the sample mean value
and sample standard deviation, respectively. The point which
generated the smallest value of the object function~(\ref{eq:obj}) in
our simulation was
\begin{eqnarray}
\theta_1 \approx 27.97^\circ, \quad \theta_2 \approx 4.45^\circ, \quad
\theta_3 \approx 43.92^\circ, \nonumber\\
\Delta m^2 \approx 2.84 \times
10^{-4} \; {\rm eV}^2, \quad \Delta M^2 \approx 1.14 \; {\rm eV}^2. \nonumber
\end{eqnarray}
Having these $20$ data points, we can now use them as initial points in a
deterministic minimization procedure. We will use a sequential
quadratic programming (SQP) method. The results obtained are
\begin{eqnarray}
&& \theta_1 = 28.7^\circ \pm 0.5^\circ, \nonumber \\
&& \theta_2 = 4.2^\circ \pm 0.1^\circ, \nonumber \\
&& \theta_3 = 45.0^\circ \pm 0.1^\circ, \nonumber \\
&& \Delta m^2 = (2.87 \pm 0.22) \times 10^{-4} \; {\rm eV}^2, \nonumber
\\
&&\Delta M^2 = 1.12 \pm 0.02 \; {\rm eV}^2 \nonumber
\end{eqnarray}
for the mean value and
\begin{eqnarray}
\theta_1 \approx 28.9^\circ, \quad \theta_2 \approx 4.2^\circ, \quad
\theta_3 \approx 45.0^\circ, \nonumber\\
\Delta m^2 \approx 2.87 \times
10^{-4} \; {\rm eV}^2, \quad \Delta M^2 \approx 1.11 \; {\rm eV}^2 \nonumber
\end{eqnarray}
for the best point.

Finally, we have evaluated the theoretically expected probabilities for
the different experiments, using the best point solution, to demonstrate
that they are indeed consistent with experimental data. This is shown
in Table~\ref{tab:probabilities}.

\begin{figure}
\begin{center}
\epsfig{figure=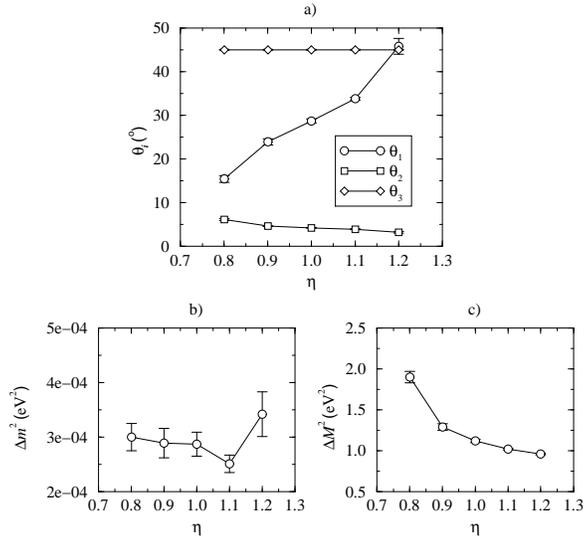,height=7cm}
\caption{The dependence on $\eta$ of the mixing angles (a) and mass
squared differences (b) and (c). For all data points except $\eta =
1.0$, we have used $N = 10$ instead of $N = 20$.}
\label{fig:eta_depend}
\end{center}
\end{figure}

Upon varying $\eta$, we have made an analysis to see the effect on the
different parameters in the model. The parameters then vary as is
shown in Fig.~\ref{fig:eta_depend}. The stability of
$\theta_{3} = 45^{\circ}$ is to be noted. The mixing angle $\theta_{2}$ varies
only slowly around $5^\circ$. The mixing angle $\theta_{1}$ varies
from $15^{\circ}$ up to $45^{\circ}$ in the interval of $\eta$
investigated. Just as expected, the mass squared differences change
slowly to compensate for this change in the atmospheric values. As
mentioned earlier, the reason for this is that the two first of the
atmospheric experiments are sensitive to the small mass squared
difference and the three others to the large mass squared
difference.

\section{Discussion and Conclusions}
\label{sec:DC}

From the results in Table~\ref{tab:probabilities}, we see that the
parameter set given as best point is indeed consistent with most of the
experiments.
It is clear from the beginning that the solar neutrino experiments
cannot all be reproduced with the parameter range used here, as was
mentioned in the Introduction. The more detailed Super-Kamiokande and
Kamiokande results, that include energy resolution of the solar
neutrinos, are very well fitted by this solution. The GALLEX
and Homestake experiments are not so well fitted. In the case of
GALLEX, the fit is within two standard deviations, whereas the
Homestake result is several standard deviations off.

For the atmospheric neutrino experiments, the MACRO and Kamiokande results are
not so well fitted. For the others, the agreement is excellent.

The accelerator neutrino experiments are all well accounted for.  It
remains to be seen if further sampling of statistics in the LSND experiment
will confirm its present value for $P_{\mu e}$.
Finally, both reactor experiments are very well reproduced by the best point
solution.

The solution to the experimental constraints all give
$\theta_{2}\approx 5^{\circ}$ and $\theta_{3} \approx 45^{\circ}$.
The mixing angle $\theta_{1}$ varies from $15^{\circ}$ to $45^{\circ}$ as
the flux factor $\eta$ varies from $0.8$ to $1.2$ times its nominal
value of 1.

Since the CKM matrix can be written as $U=U_{23}(\theta_{1})
U_{13}(\theta_{2})U_{12}(\theta_{3})$, where $U_{ij}(\theta)$ is a
rotation by an angle $\theta$ in the $ij$ plane, it is clear that
$\theta_{3}=45^{\circ}$
corresponds to maximal mixing between the two lightest neutrinos,
$\nu_{1}$ and $\nu_{2}$.
There are two particularly simple ``solutions'' more or less in the
ranges of the mixing angles: (a) bimaximal mixing \cite{barg982} with
$\theta_{1} = 45^{\circ}$,
$\theta_{2} = 0$, and $\theta_{3} = 45^{\circ}$ and (b) single maximal mixing
with $\theta_{1} = 30^{\circ}$, $\theta_{2} = 0$, and $\theta_{3} =
45^{\circ}$.
These two ``solutions'' both reproduce data quite well (apart from
the LSND experiment) the atmospheric ones with the appropriate factor
$\eta$ included. The CKM matrices for the simple ``solutions'' are given by
$$
U_a = \left( \begin{array}{ccc} \frac{1}{\sqrt{2}} &
\frac{1}{\sqrt{2}} & 0 \\ - \frac{\sqrt{3}}{2 \sqrt{2}} &
\frac{\sqrt{3}}{2 \sqrt{2}} & \frac{1}{2} \\ \frac{1}{2 \sqrt{2}} & -
\frac{1}{2 \sqrt{2}} & \frac{\sqrt{3}}{2} \end{array} \right)
$$
and
$$
U_b = \left( \begin{array}{ccc} \frac{1}{\sqrt{2}} &
\frac{1}{\sqrt{2}} & 0 \\ - \frac{1}{2} &
\frac{1}{2} & \frac{1}{\sqrt{2}} \\ \frac{1}{2} & -
\frac{1}{2} & \frac{1}{\sqrt{2}} \end{array} \right)
$$
to be compared with the CKM matrix for the best point solution
$$
U \simeq \left( \begin{array}{ccc} 0.7052 & 0.7052 & 0.0732 \\ -0.6441
& 0.5940 & 0.4820 \\ 0.2964 & -0.3871 & 0.8731 \end{array} \right).
$$
Note that the matrix element $U_{e3}$ is small compared to all the other matrix
elements, in agreement with earlier suggestions \cite{viss97}. For the simple
``solutions,'' it becomes identically equal to zero.

\begin{figure}
\begin{center}
\epsfig{figure=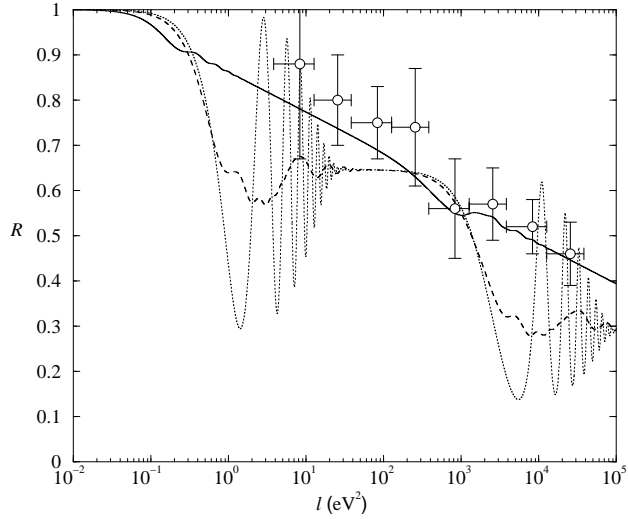,height=7cm}
\caption{The ``ratio of ratios'' ${\cal R}$ as a function of $l
\simeq 1.27 L/E$. The data points are
obtained from Fig.~4 in Ref. \protect\cite{fuku983}. The dotted
curve is a theoretical curve using the best point solution. The dashed
and solid curves are obtained from the theoretical curve by using
running average with different averaging lengths. The solid curve uses
a larger averaging length than the dashed curve.}
\label{fig:R_l}
\end{center}
\end{figure}

For the double ratio, we obtain approximately
\begin{equation}
{\cal R} \equiv R_{\mu}/R_{e} \simeq \frac{1}{R} \frac{R P_{\mu\mu} +
P_{e\mu}}{P_{ee} + RP_{\mu e}}.
\label{eq:ratio}
\end{equation}
In our case, the CKM matrix is real so $P_{e \mu}=P_{\mu e}$. Using the
published \cite{hond95} values for the theoretical
$\nu_{e}/\nu_{\mu}$ flux ratio for the upward-going muon-neutrinos
$(\cos \vartheta \leq -0.6)$, we have calculated the ratio of ratios
for the solution obtained with $\eta=1$. This ratio also eliminates
the uncertainty in the muon-neutrino flux from the atmosphere. The
result is ${\cal R} = 0.51 \pm 0.02$, in good agreement with the value
given by the Super-Kamiokande Collaboration of ${\cal R} = 0.41 \pm
0.12$ for multi-GeV data. This indicates that the electron-neutrino
ratio $R_e$ is also in agreement with data. See Fig.~\ref{fig:R_l} for
the $L/E$ dependence of ${\cal R}$.

Our solution is also consistent with the $({\rm UP}/{\rm
DOWN})_{\mu}$ asymmetry measured by the Super-Kamiokande Collaboration
\cite{kaji98}, as well as with the corresponding $({\rm UP}/{\rm
DOWN})_{e}$ asymmetry. With our approximations, the $({\rm UP}/{\rm
DOWN})_{\mu}$ asymmetry is given by
\begin{equation}
U_\mu/D_\mu \simeq P_{\mu\mu} + \frac{1}{R} P_{e\mu} \approx P_{\mu\mu},
\label{updown}
\end{equation}
where $U_\mu$ is the number of $\mu$-like upward-going events and
$D_\mu$ is the number of $\mu$-like downward-going events.
The measured value for multi-GeV events is $U_\mu/D_\mu = 0.54 \pm 0.07$
\cite{kaji98}, which is consistent with the value $P_{\mu\mu} = 0.56
\pm 0.07$ obtained from our analysis when $\eta = 1$.

In the present analysis, we have deliberately considered mass ranges for the
solution that avoid the region where the solar (and atmospheric) neutrinos
could be affected by the MSW effect. We plan to address
this question specifically in a future analysis. The mass range for
the solar neutrino problem will then interfere with the analysis of
atmospheric neutrinos.

\acknowledgments

This work was supported by the Swedish Natural Science Research
Council (NFR), Contract No. F-AA/FU03281-312. Support for this work
was also provided by the Engineer Ernst Johnson Foundation (T.O.).
T.O. would like to thank Samoil M. Bilenky, Manfred Lindner, and Martin Freund
for stimulating discussions on neutrino physics.

\newpage

\onecolumn

\squeezetable

\begin{table}
\caption{Neutrino experiments. The abbreviations SBL and LBL stand for
short-baseline and long-baseline, respectively.}
\begin{tabular}{lllcccc}
Experiment & Type & Reaction & $L$ (m) & $E$ (MeV) & $E/L$ (${\rm
eV}^2$) & $P_{\rm exp}$ \\
\hline
SAGE \cite{gavr98} & solar & $\nu_e \to \nu_e$ & $1.496 \times 10^{11}$
& $\approx 1.1$ & $10^{-12}$--$10^{-10}$ & $0.52 \pm 0.06$
\tablenotemark[1] \\
GALLEX \cite{anse95} & solar & $\nu_e \to \nu_e$ & $1.496 \times
10^{11}$ & $\approx 1.1$ & $10^{-12}$--$10^{-10}$ & $0.60 \pm 0.06$
\tablenotemark[1] \\
Homestake \cite{clev98} & solar & $\nu_e \to \nu_e$ & $1.496 \times 10^{11}$ &
$\approx 5.5$ & $10^{-11}$--$10^{-10}$ & $0.33 \pm 0.029$
\tablenotemark[1] \\
Super-Kamiokande \cite{fuku98} & solar & $\nu_e \to \nu_e$ & $1.496
\times 10^{11}$ & $\approx 10$ & $10^{-11}$--$10^{-10}$ & $0.474 \pm 0.020$
\tablenotemark[1] \\
Kamiokande \cite{fuku96} & solar & $\nu_e \to \nu_e$ & $1.496 \times 10^{11}$ &
$\approx 10$ & $10^{-11}$--$10^{-10}$ & $0.54 \pm 0.07$
\tablenotemark[1] \\
\hline
Soudan \cite{alli97} & atmospheric & $\nu_\mu \to \nu_\mu$ & $1.03
\times 10^7$ &
$\approx 1~000$ & $\sim 10^{-4}$ & $\sim 0.45 \pm 0.15$ \cite{gall98} \\
Super-Kamiokande \cite{kaji98,fuku982,fuku983} & atmospheric &
$\nu_\mu \to \nu_\mu$ & $1.03 \times 10^7$ & $\approx 10~000$ & $\sim
10^{-3}$ & $0.56 \pm 0.07$ \cite{kaji98,saka99} \\
MACRO \cite{ahle95} & atmospheric & $\nu_\mu \to \nu_\mu$ & $1.22
\times 10^7$ &
$\approx 100~000$ & $\sim 10^{-2}$ & $\sim 0.5 \pm 0.1$ \cite{ambr98} \\
Kamiokande \cite{kami} & atmospheric & $\nu_\mu \to \nu_\mu$ & $1.22
\times 10^7$ & $\approx 100~000$ & $\sim 10^{-2}$ & $\sim 0.85 \pm 0.15$
\cite{hata98} \\
IMB \cite{casp91} & atmospheric & $\nu_\mu \to \nu_\mu$ & $1.03 \times 10^7$ &
$\approx 250~000$ & $10^{-1}$--$10^2$ & $\sim 0.7 \pm 0.1$
\cite{beck92} \\
\hline
LSND \cite{atha95} & accelerator (SBL) & $\nu_\mu \to \nu_e$ & $30$ &
$48$ & $1$--$2$ & $0.0026 \pm 0.0010 \pm 0.0005$ \\
KARMEN \cite{zeit98} & accelerator (SBL) & $\nu_\mu \to \nu_e$ &
$17.7$ & $\approx 40$ & $1$--$4$ & $\leq 0.0031$ \\
NOMAD \cite{alte98} & accelerator (SBL) & $\nu_\mu \to \nu_\tau$ &
$625$ & $24000$ & $\sim 10$ & $\leq 0.0021$ \\
CHORUS \cite{migl98} & accelerator (SBL) & $\nu_\mu \to \nu_\tau$ &
$600$ & $26620$ & $10$--$100$ & $\leq 0.0006$ \\
\hline
CHOOZ \cite{apol98} & reactor (LBL) & $\nu_e \to \nu_e$ & $1030$ & $3$ & $\sim
10^{-3}$ & $0.98 \pm 0.04 \pm 0.04$ \\
Bugey \cite{achk95} & reactor (SBL) & $\nu_e \to \nu_e$ & $40$ & $3$ &
$10^{-2}$--$10^{-1}$ & $0.99 \pm 0.01 \pm 0.05$ \\
\end{tabular}
\tablenotetext[1]{Normalized according to the Bahcall-Basu-Pinsonneault
1998 standard solar model (BBP98 SSM) \cite{bahc98,bahc982}.}
\label{tab:experiments}
\end{table}

\begin{table}
\caption{Probabilities. The best point solution obtained in our
simulation, using first stochastic (s.) and then deterministic (d.)
minimization, is compared to other ``solutions.'' For comparison of
the different solutions, the $\chi^2$ values have also been listed.}
\begin{tabular}{lcccc}
Experiment & $P_{\rm exp}$ & $P$ (best; s. + d.) & $P$
\tablenotemark[1] & $P$ \tablenotemark[2] \\
\hline
SAGE & $0.52 \pm 0.06$ & 0.49 & 0.50 & 0.50 \\
GALLEX & $0.60 \pm 0.06$ & 0.49 & 0.50 & 0.50 \\
Homestake & $0.33 \pm 0.029$ & 0.49 & 0.50 & 0.50 \\
Super-Kamiokande & $0.474 \pm 0.020$ & 0.49 & 0.50 & 0.50 \\
Kamiokande & $0.54 \pm 0.07$ & 0.49 & 0.50 & 0.50 \\
\hline
Soudan & $\sim 0.45 \pm 0.15$ & 0.45 & 0.27 & 0.35 \\
Super-Kamiokande & $0.56 \pm 0.07$ & 0.56 & 0.45 & 0.54 \\
MACRO & $\sim 0.5 \pm 0.1$ & 0.64 & 0.50 & 0.62 \\
Kamiokande & $\sim 0.85 \pm 0.15$ & 0.64 & 0.50 & 0.62 \\
IMB & $\sim 0.7 \pm 0.1$ & 0.64 & 0.50 & 0.62 \\
\hline
LSND & $0.0026 \pm 0.0015$ & 0.0030 & $3.8 \times 10^{-8}$ &
$4.2 \times 10^{-8}$ \\
KARMEN & $\leq 0.0031$ & 0.0017 & $1.9 \times 10^{-8}$ & $2.1 \times
10^{-8}$ \\
NOMAD & $\leq 0.0021$ & $9.5 \times 10^{-4}$ & $1.1 \times 10^{-3}$ &
$9.9 \times 10^{-4}$ \\
CHORUS & $\leq 0.0006$ & $7.1 \times 10^{-4}$ & $8.2 \times 10^{-4}$ &
$7.4 \times 10^{-4}$ \\
\hline
CHOOZ & $0.98 \pm 0.08$ & 0.97 & 0.98 & 0.98 \\
Bugey & $0.99 \pm 0.06$ & 0.99 & 1.00 & 1.00 \\
\hline
$\chi^2 = F$ & & $3.48 \times 10^{-2}$ & $6.22 \times 10^{-2}$ & $6.14
\times 10^{-2}$ \\
\end{tabular}
\tablenotetext[1]{Parameter values: $\theta_1 = 45^\circ$, $\theta_2 =
0$, $\theta_3 = 45^\circ$, $\Delta m^2 = 3.5 \times 10^{-4} \; {\rm
eV}^2$, $\Delta M^2 = 1 \; {\rm eV}^2$, and $\eta = 1.2$.}
\tablenotetext[2]{Parameter values: $\theta_1 = 30^\circ$, $\theta_2 =
0$, $\theta_3 = 45^\circ$, $\Delta m^2 = 3 \times 10^{-4} \; {\rm
eV}^2$, $\Delta M^2 = 1.1 \; {\rm eV}^2$, and $\eta = 1.0$.}
\label{tab:probabilities}
\end{table}

\end{document}